\documentclass[journal,10pt,a4paper,twocolumn,twoside]{IEEEtran}

\usepackage{graphicx}

\usepackage{subcaption}
\usepackage{amsmath}
\usepackage{booktabs}

\usepackage{xcolor}
\usepackage{amssymb}
\usepackage{amsmath,cases}
\usepackage{dsfont}
\usepackage{physics}

\usepackage{hyperref}
\usepackage{cite}
\usepackage{newpxtext,newpxmath}
\usepackage{acronym}

\definecolor{ocre}{HTML}{800000}
\definecolor{green}{RGB}{0, 128, 0}
\definecolor{sky}{HTML}{C6D9F1}
\definecolor{skybox}{HTML}{5F86B3}

\usepackage{tikz}
\usetikzlibrary{quantikz}
\usetikzlibrary{decorations.text, arrows.meta,calc,shadows.blur,shadings,intersections}
\usetikzlibrary{trees}

\begin{document}

\title{Spatial-Mode Diversity and Multiplexing for Continuous Variables Quantum Communications}

\author{Seid Koudia*\thanks{*The corresponding authors seid.koudia@uni.lu, symeon.chatzinotas@uni.lu}, Leonardo Oleynik, Mert Bayraktar, Junaid ur Rehman, and Symeon Chatzinotas*,~\IEEEmembership{Fellow,~IEEE}\\
	
	\thanks{
 The authors are with the Interdisciplinary Centre for Security, Reliability, and Trust (SnT), Luxembourg, L-1855 Luxembourg. 

 }
    }

\maketitle

\begin{abstract}
We investigate the performance of continuous-variable (CV) quantum communication systems employing diversity schemes to mitigate the effects of realistic channel conditions, including Gaussian lossy channels, fading, and crosstalk. By modeling the transmittivity of the channel as a log-normal distribution, we account for the stochastic nature of fading. We analyze the impact of both post-processing amplification at the receiver and pre-amplification at the transmitter on the fidelity of the communication system. Our findings reveal that diversity schemes provide significant advantages over single-channel transmission in terms of fidelity and secret key rate, particularly in conditions of strong fading and high thermal background noise. We also explore the effect of crosstalk between channels and demonstrate that a noticeable advantage persists in scenarios of strong fading or thermal noise. For CV-QKD, we show that diversity can outperform multiplexing revealing a a diversity advantage over multiplexing in some regimes.  
\end{abstract}


\section{Introduction}
Continuous Variable (CV) quantum communications have emerged as a powerful and versatile approach to secure information transfer \cite{braunstein2005quantum}, encompassing both Quantum Key Distribution (QKD) and entanglement-based communications \cite{zhang2024continuous}. Unlike discrete variable systems, which rely on single photons, CV quantum communications utilize the continuous properties of light, such as amplitude and phase quadratures, offering distinct advantages in terms of implementation and integration with existing telecommunications infrastructure \cite{andersen2010continuous}.

In the realm of Quantum Key Distribution, CV-QKD protocols provide a robust framework for generating and distributing cryptographic keys with unconditional security \cite{PhysRevLett.110.030502}, rooted in the principles of quantum mechanics. These protocols leverage Gaussian-modulated coherent states and homodyne or heterodyne detection techniques \cite{jouguet2011long}, making them compatible with standard optical components and allowing for potentially higher key rates compared to their discrete variable counterparts \cite{jouguet2013experimental}. Furthermore, CV-QKD systems can operate at room temperature and do not require the use of single-photon detectors, which simplifies their deployment and reduces costs \cite{koudia2024physical}.

Entanglement-based communications \cite{koudia2023quantum}, on the other hand, exploit the fundamental properties of quantum entanglement to enable applications such as quantum teleportation, dense coding, and distributed quantum computing. CV entanglement offers the advantage of generating and manipulating entangled states with relatively high efficiency and fidelity, facilitating the creation of long-distance quantum networks \cite{dias2020quantum}.

One of the key challenges in CV quantum communications is the inherent sensitivity to noise and losses, which can significantly degrade the performance of quantum protocols. To address these issues and enhance the reliability and efficiency of CV quantum communication systems, two pivotal strategies can be identified: multiplexing and diversity.

Multiplexing techniques \cite{koudia2024physical}, including wavelength-division multiplexing (WDM) \cite{choi2010quantum}, mode-division multiplexing (MDM) \cite{zhou2021multiprobe} and Spatial multiplexing (SM) \cite{xavier2020quantum}, enable the simultaneous transmission of multiple quantum channels over  multicore/multimode fibers and free space links affected by crosstalk \cite{eriksson2019crosstalk}. This approach not only increases the overall communication rates but also optimizes the utilization of available bandwidth, thereby enhancing the scalability of quantum networks. By integrating multiplexing methods, it is possible to achieve higher data throughput and more efficient use of resources, which is crucial for the practical deployment of large-scale quantum communication systems.

Diversity, on the other hand, plays a critical role in mitigating the effects of noise and losses, which are inevitable in real-world communication channels. Spatial diversity and temporal diversity are among the techniques employed to enhance the robustness of classical communications, and being recently studied in the quantum realm \cite{wang2024exploiting,yuan2020free,oleynik2024diversity,anipeddi2024optical}. By transmitting quantum information through multiple independent paths or using different degrees of freedom, it is possible to reduce the likelihood of complete signal degradation, thereby improving the resilience of the communication system against environmental perturbations and adversarial attacks. Nevertheless, due to the no-cloning theorem, these methods cannot be applied to unknown quantum states, as they cannot be copied. Diversity based on asymmetric cloning has been proposed to tackle this problem \cite{oleynik2024diversity}. 

In this paper, diversity schemes, involving multiple transmission paths and varying channel parameters, have been proposed to enhance the robustness of quantum communication systems.  We examine the effectiveness of such schemes in CV quantum communications by considering the impact of fading, modeled as a log-normal distribution, and crosstalk. We quantify the performance improvements in terms of fidelity and secret key rate, an essential metric for the evaluation of CV-QKD systems. The paper is structured as follows. In Sec.~\ref{sec:02}, we present the different models adopted in this work including the channel model, the crosstalk model and Gaussian ampification, as well as the figures of merit used for benchmarking the results, namely, fidelity and the secret key rate. In Sec.~\ref{sec:03} we present the main results of the paper. For instance, we discuss the proposed diversity scheme and its advantage using different amplifications depending on the CSI available to the transceivers. We discuss the effect of crosstalk on the achieved average fidelity as well as the advanatge of the diversity scheme for CV QKD in terms of the average secret key rate. Sec.~\ref{sec:04} concludes the paper.
\section{Model}
\begin{figure*}
    \begin{minipage}[c] {0.3\textwidth}
        \centering
        \includegraphics[width=1\columnwidth]{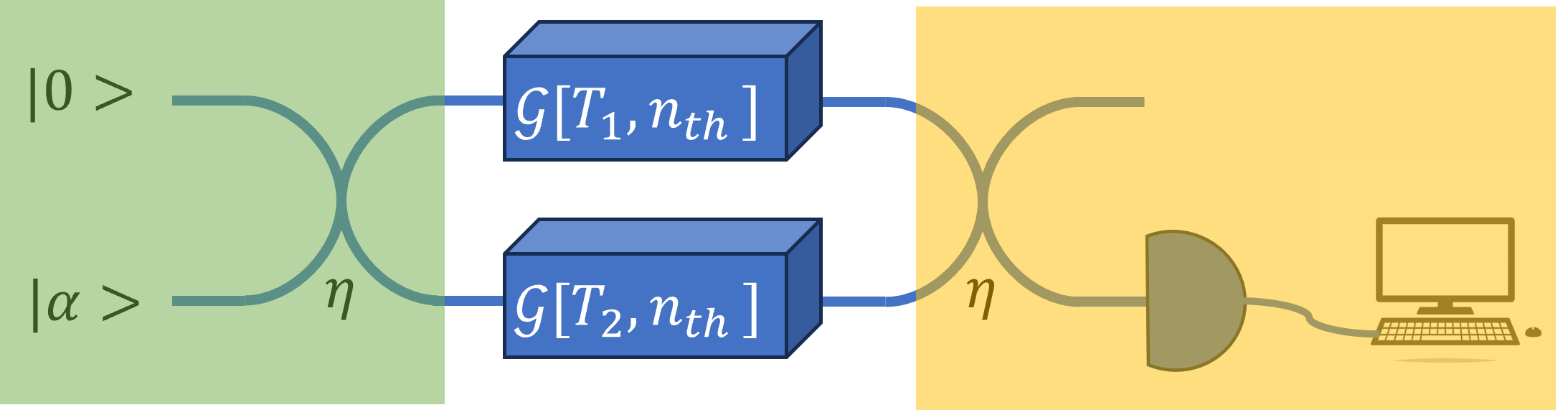}
        \subcaption{A diversity scheme with postprocessing amplification of the signal upon CSI at the receiver}
        \label{Fig:04-a}
    \end{minipage}
    \hspace{0.02\textwidth}
   \begin{minipage}[c] {0.3\textwidth}
        \centering
        \includegraphics[width=1\columnwidth]{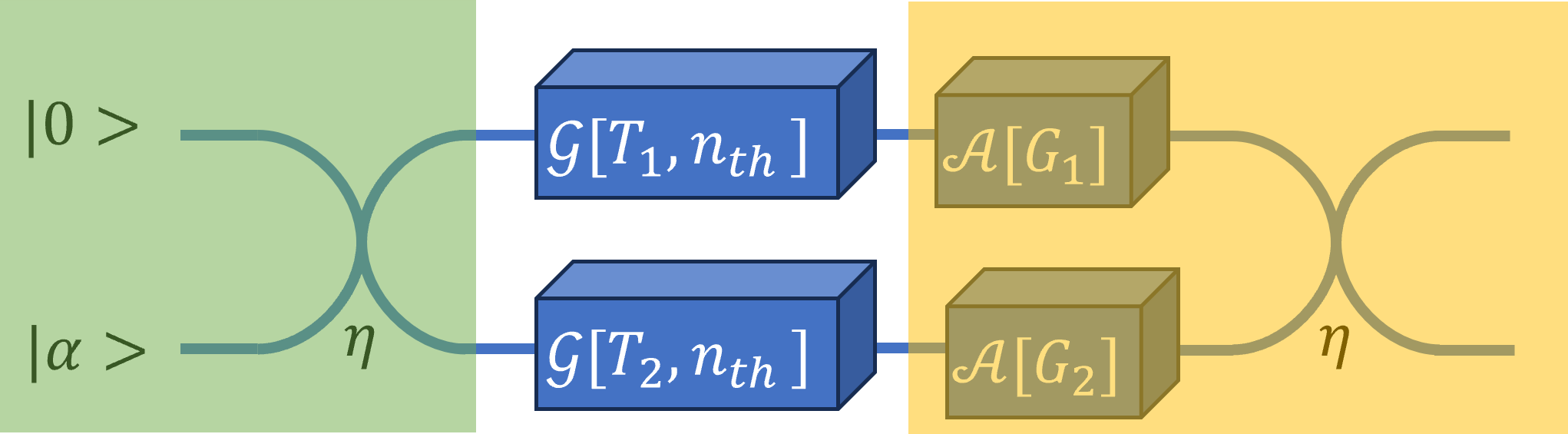}
        \subcaption{A diversity scheme with postamplification of the signal upon CSI at the receiver}
        \label{Fig:04-b}
    \end{minipage}
    \hspace{0.02\textwidth}
      \begin{minipage}[c] {0.3\textwidth}
        \centering
        \includegraphics[width=1\columnwidth]{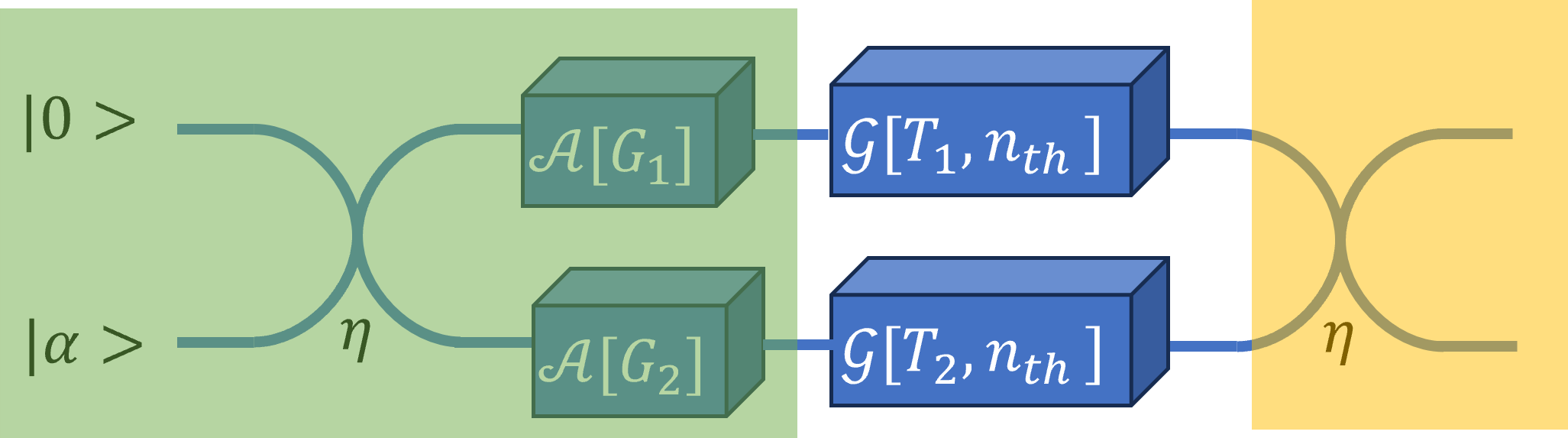}
        \subcaption{A diversity scheme with preamplification of the signal with CSI at the transmitter.}
        \label{Fig:04-c}
    \end{minipage}
    \caption{A diversity scheme with different amplification techniques with green and yellow shadowed boxes referring to the transmitter and the receiver respectively.}
    \label{fig:diversity}
    \hrulefill
\end{figure*}
\label{sec:02}
\subsection{Channel Model}
We consider Gaussian Lossy channels $\mathcal{G}[T,n_{th}]$, described by a transmittivity $T$ and thermal noise characterized by the thermal background average number of photons $n_{th}$. The channel acts on a Gaussian state $\rho$ of covariance matrix $V$ and mean vector $d$ as:
\begin{align}
\label{gaussian channel}
    \rho & \rightarrow  \mathcal{G}[T,n_{th}] (\rho) \nonumber\\
    V &\rightarrow T V+ (1-T) (n_{th}+\frac{1}{2})\mathrm{I}_2 \nonumber\\
    d & \rightarrow \sqrt{T}d
\end{align}
with $\mathrm{I}_2$ being the two dimensional identity matrix. 

The transmitivity  being stochastic due to fading, is modeled by a log-normal distribution 
\begin{equation}
    f_T(T) = \frac{1}{T \sigma_T \sqrt{2\pi}} \exp\left(-\frac{(\ln T - \mu_T)^2}{2\sigma_T^2}\right)
\end{equation}
with
the mean $\mu_T$ reflecting the average behaviour of transmittivity of the channel and  the variance $\sigma_T$ quantifying the fading strength of the channel. 

\subsection{Gaussian amplification}
The process of a physical active amplification $\mathcal{A}[G]$ of the signal with amplification gain $G$ is described by a Gaussian channel that transforms the original state by:
\begin{align}
    \rho & \rightarrow  \mathcal{A}[G] (\rho)\\
    V &\rightarrow G V+ (G-1)\mathrm{I}_2\\
    d & \rightarrow \sqrt{G}d
\end{align}
The term $(G-1)\mathrm{I}_2$ reflects that the active physical amplification is a noisy process.
\subsection{Crosstalk}
Crosstalk in communications can arise in various scenarios and is typically due to interference between signal paths.  As the optical signal propagates through free space, the light beam can diverge due to diffraction. If the beams from different transmitters are not sufficiently collimated or if they diverge too much, they can overlap at the receiver, causing crosstalk. Moreover, variations in the refractive index of the atmosphere caused by temperature fluctuations and air movement can lead to beam wandering and spreading.  Similaraly, crosstalk in multicore fibers (MCFs) primarily arises from core-to-core coupling due to the close proximity of the cores, where overlapping evanescent fields cause optical modes in one core to couple into adjacent cores. This issue is exacerbated by refractive index fluctuations, geometry non-uniformities, and mechanical stresses such as fiber bending and external forces. Additionally, propagation effects like differential mode delay and random perturbations can dynamically influence crosstalk levels. Crosstalk can also vary with wavelength and modal characteristics. 
Crosstalk can cause parts of the beam to stray into the path of another receiver.
The crosstalk in this paper is modalized as a beam splitter  $B(\eta)$  with $\eta$ being its  transmittivity. The $B$ representation on phase space is given by the symplectic matrix: 
\begin{equation}
   B(\eta)= \begin{pmatrix}
        \sqrt{\eta} \mathrm{I}_2& -\sqrt{1-\eta} \mathrm{I}_2\\
         \sqrt{1-\eta} \mathrm{I}_2& \sqrt{\eta} \mathrm{I}_2
    \end{pmatrix}
\end{equation}
The action of $B$ on the covariance  matrix $V$ and the mean vector $d$ of a given state is described as:
\begin{align}
    V & \rightarrow B(\eta)VB(\eta)^T\\
    d &\rightarrow B(\eta)d
\end{align}
\subsection{Figure of Merit}
\subsubsection{Fidelity}
We benchmark the advantage of the scheme in terms of the average fidelity in the 2-diversity and the 0-diversity scenarios, where the latter denotes single transmission. The fidelity between the output state and input state with Wigner functions $W_{out}(\chi)$ and $W_{in}(\chi)$ as:
\begin{equation}
    F(\rho_{out},\rho_{in})=(2\pi)^n \int W_{out}(\chi)W_{in}(\chi)d\chi
\end{equation}
where $n$ is the number of modes of the states. For two Gaussian states, $\rho_{in}$ and $\rho_{out}$ of phase space representation $(V1,d1)$ and $(V2,d2)$ respectively, the fidelity expression simplifies to:
\begin{equation}
    F(\rho_{out},\rho_{in})=\frac{\exp\Big(-\frac{(d1-d2) (V1+V2) (d1-d2))^T}{2}\Big)}{\sqrt{\det(V1+V2)}}
\end{equation}
In a realistic communications scenario, the mean vectors and the covariance matrices of the output states $\rho_{out}$ depends on the stochastic transmittivity parameter. As a mater of fact, we define the average fidelity between the output and inuut states by:
\begin{equation}
    F_{avg}=\int f_T(T)F(\rho_{out}(T),\rho_{in}(T))dT
\end{equation}
\subsubsection{Secret Key rate}
The secret key rate is a crucial figure of merit in the evaluation of diversity schemes in quantum key distribution (QKD). It quantifies the amount of secure key bits generated per unit time or per channel use, directly reflecting the efficiency and security of the QKD system. In practical implementations, especially in diverse and dynamic environments such as free-space optical communication or fiber networks with varying conditions, the secret key rate becomes even more significant. Diversity schemes, which involve using multiple transmission paths or varying channel parameters to mitigate the effects of noise, loss, and eavesdropping, aim to maximize the secret key rate under these fluctuating conditions. By integrating the impact of different channel characteristics, such as lognormally distributed transmittivity, the secret key rate provides a comprehensive measure of the QKD system's resilience and performance. It allows for the comparison of different diversity strategies, ensuring that the chosen approach not only maintains high security but also optimizes key generation efficiency across varying scenarios, thereby enhancing the overall robustness of the quantum communication network.
For a continuous-variable QKD (CV-QKD) system using Gaussian modulation with coherent states, the averagesecret key rate $K$ with direct reconcilliation can be expressed as \cite{laudenbach2018continuous}:
\begin{equation}
    K=\int f_T(T)\Big(\beta I(A:B)-\chi(E:B)\Big)dT
\end{equation}
where $\beta$ is the reconciliation efficiency, $I(A:B)$ is the mutual information between the input and output states and $\chi(E:B)$ is the Holevo information between the environment and the ouput given by:
\begin{equation}
   \chi(E:B)= S(E)-S(E|B)
\end{equation}
With $S(E)$ being the entropy of the reduced state of the environment, which is holding the purification of the system, and $S(E|B)$ is the entropy of the environment state conditioned on the outcome of the measurement of the receiver $B$. The entropy function is given by:
\begin{equation}
    S(x)=(x+1)\log_2(x+1)-x\log_2(x)
\end{equation}

\section{Results}
\label{sec:03}
Depending on the Channel state information (CSI) available to the transceivers, we distinguish three main scenarios for diversity. When the receiver has CSI about the average statistical behaviour of the Gaussian lossy channel, namely knowing $\langle T \rangle$, he may apply an amplification technique to overcome the losses during transmission. He might apply active amplification (we limit ourselves to phase insensitive amplification) by interacting with the system before any measurement, or he may carry postprocessing amplification after having some measurement outcome statistics. In the contrary, if the transmitter has CSI about the average behaviour of the Gaussian lossy channel, he may apply some pre-amplification before transmitting the signal in order to shield the original signal from losses. These different schemes are illustrated in Fig.~\ref{fig:diversity}

\subsection{Postprocessing Amplification}
\begin{figure}
    \centering
    \includegraphics[width=1\linewidth]{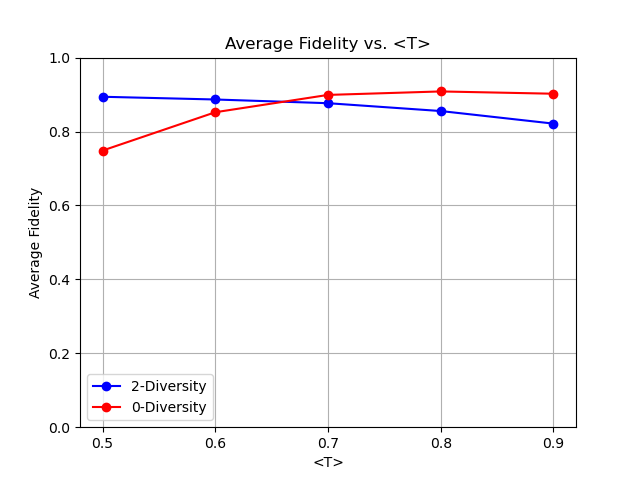}
    \caption{The average fidelity in the 2-diversity and the 0-diversity schemes as a function of the average transmittivity of the fading channel for $n_{th}=0.9$ for the thermal Gaussian noise }
    \label{fig:1}
\end{figure}

\begin{figure}
    \centering
    \includegraphics[width=1\linewidth]{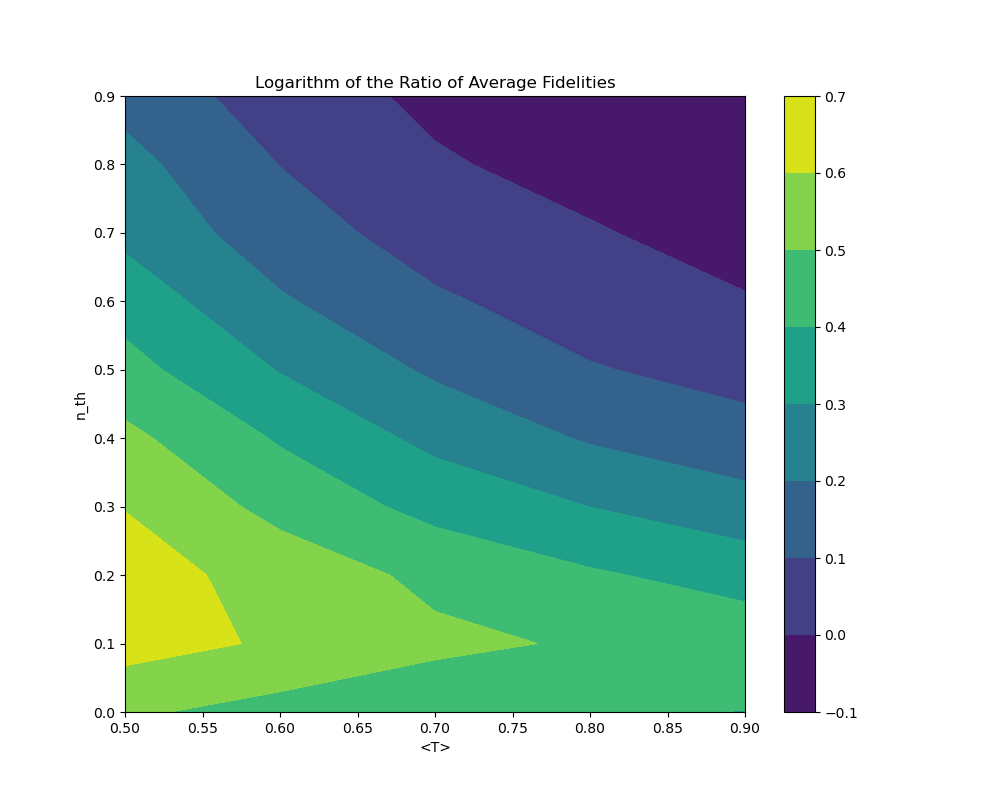}
    \caption{The log scale of the ratio between the average fidelities in the 2-diversity and the 0-diversity schemes as a function of the average transmitivitty of the fading channels and the strength of the thermal Gaussian noise.}
    \label{fig:02}
\end{figure}
We assume that the Receiver has a CSI regarding the average behaviour of the fading channel. As such, he might apply post-processing amplification after performing a heterodyne measurement on the quadratures of the received signal. This is valid when transmitting classical information encoded in the quadratures of the Gaussian state. This postprocessing amplification does not interact with signal actively, hence, it does not introduce any additional noise or corrections to the covariance matrix, altough, it does only rescale the output state's quadratures
\begin{equation}
    \Vec{d}\rightarrow G \Vec{d}
\end{equation}
where $G$ is the postprocessing amplification gain.
The global evolution of the system is given by:
\begin{align}
    V_1\oplus V_2 &\rightarrow B(\eta)\Big[(K_1\oplus K_2)(V_1\oplus V_2) (K_1\oplus K_2)^T \nonumber\\
    &+ (N_1\oplus N_2)\Big]B(\eta)^T \nonumber\\
    d_1\oplus d_2 &\rightarrow G \cdot B(\eta) (K_1\oplus K_2) (d_1\oplus d_2)
\end{align}
The covariance matrix of the states evolve after the described process as follows:
\begin{align}
 V^{(2)}_{in}=\begin{pmatrix}
         \frac{\mathrm{I}_2}{2}& 0\\
         0& \frac{\mathrm{I}_2}{2}
    \end{pmatrix} \rightarrow  V^{(2)}_{out}=\begin{pmatrix}
         V1& V2\\
         V2& V3
    \end{pmatrix} 
\end{align}
where 
\begin{align}
    V1&=(\eta B_1+(1-\eta)B_2)\mathrm{I}_2\\
    V2&=(\sqrt{\eta (1-\eta)}(B_1-B_2))\mathrm{I}_2\\
    V_3&=(\eta B_2+(1-\eta)B_1)\mathrm{I}_2    
\end{align}
 with 
 \begin{align}
     B_1&=T_1(\frac{1}{2}-(n_{th}+\frac{1}{2}))+(n_{th}+\frac{1}{2})\\
     B_2&=T_2(\frac{1}{2}-(n_{th}+\frac{1}{2}))+(n_{th}+\frac{1}{2})
 \end{align}
 The mean vectors evolve according to the described processs as:
 \begin{align}
     d_1=0&\rightarrow \Bigg[\Bigg(\sqrt{\eta T_1}\frac{x}{\sqrt{2\langle T_1\rangle}}-\sqrt{(1-\eta) T_2}\frac{x}{\sqrt{2\langle T_1\rangle}}\Bigg) \nonumber\\
     &+ i \Bigg(\sqrt{\eta T_1}\frac{p}{\sqrt{2\langle T_1\rangle}}-\sqrt{(1-\eta) T_2}\frac{p}{\sqrt{2\langle T_1\rangle}}\Bigg)\Bigg]\nonumber\\
     d_2=x+ip&\rightarrow  \Bigg[\Bigg(\sqrt{\eta T_2}\frac{x}{\sqrt{2\langle T_2\rangle}}+\sqrt{(1-\eta) T_1}\frac{x}{\sqrt{2\langle T_2\rangle}}\Bigg) \nonumber\\
     &+ i \Bigg(\sqrt{\eta T_2}\frac{p}{\sqrt{2\langle T_2\rangle}}+\sqrt{(1-\eta) T_1}\frac{p}{\sqrt{2\langle T_2\rangle}}\Bigg) \Bigg] 
 \end{align}
The results for fixed thermal background noise to $n_{th}=0.9$ are shown in Fig.~\ref{fig:1}. We notice that a 2-diversity scheme is performing better than single transmission in very strong fading channels. We stress that the advantage depends also on the thermal background noise. This dependence is highlighted in Fig.~\ref{fig:02}.

\subsection{Post-Amplification}
\begin{figure}
    \centering
    \includegraphics[width=1\linewidth]{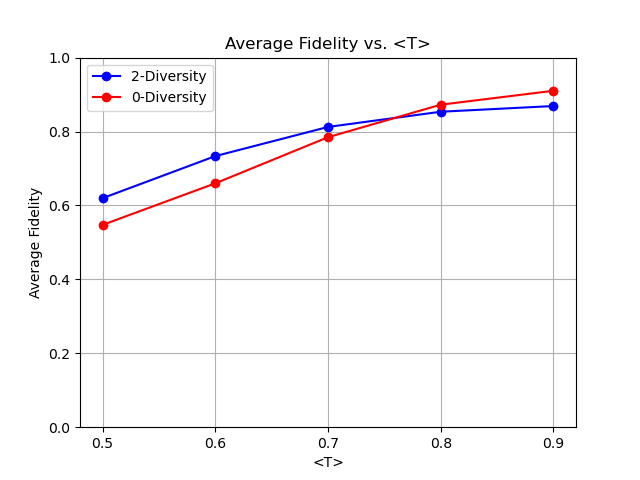}
    \caption{Post-Amplification: The average fidelity in the 2-diversity and the 0-diversity schemes as a function of the average transmittivity of the fading channel for $n_{th}=0.7$ for the thermal Gaussian noise }
    \label{fig:post_average}
\end{figure}

\begin{figure}
    \centering
    \includegraphics[width=1\linewidth]{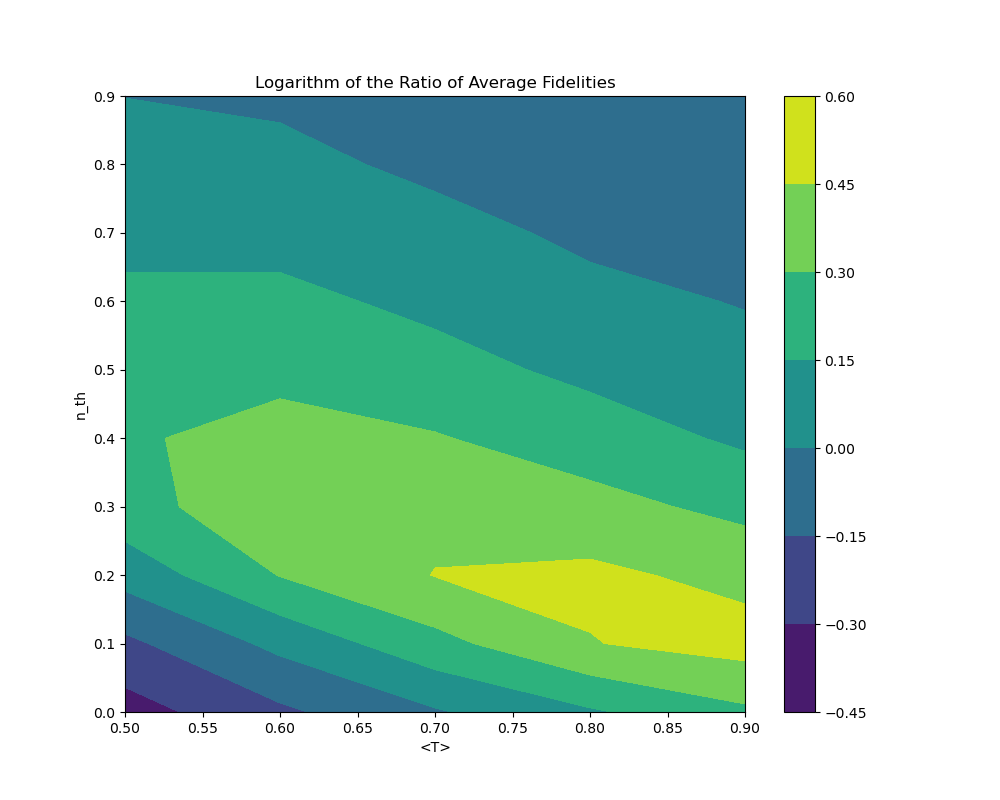}
    \caption{The log scale of the ratio between the average fidelities in the 2-diversity and the 0-diversity schemes as a function of the average transmitivitty of the fading channels and the strength of the thermal Gaussian noise.}
    \label{fig:post_contour}
\end{figure}
We assume that CSI about the average behaviour of the fading is available to the receiver. Accordingly, he makes a post-amplification to overcome the effect of losses. This postamplification mechanism is different from the postprocessing mechanism as this one is allowing the receiver to actively interact with the signal. Therefore, an additional noise to the covariance of the signal is unavoidable. Indeed, the strength of the fading can be different in the two channels, hence different amplification gains should be carried depending on the individual CSI.
The global evolution of the system is given by:
\begin{align}
    &V_1\oplus V_2 \rightarrow\nonumber\\
    & B(\eta)\Bigg[(A(G_1)\oplus A(G_2))\Big[(K_1\oplus K_2)(V_1\oplus V_2) (K_1\oplus K_2)^T \nonumber\\
    &+ (N_1\oplus N_2)\Big](A(G_1)\oplus A(G_2))^T + N_{A(G_1)}\oplus N_{A(G_2)}\Bigg] B(\eta)^T \nonumber\\
    &d_1\oplus d_2 \rightarrow  B(\eta)  (A(G_1)\oplus A(G_2))(K_1\oplus K_2)(d_1\oplus d_2)
\end{align}
Explicitly, the covariance matrix evolves as:
\begin{align}
 V^{(2)}_{in}=\begin{pmatrix}
         \frac{\mathrm{I}_2}{2}& 0 \nonumber\\
         0& \frac{\mathrm{I}_2}{2}
    \end{pmatrix} \rightarrow  V^{(2)}_{out}=\begin{pmatrix}
         V1& V2\\
         V2& V3
    \end{pmatrix} 
\end{align}
\begin{align}
    V1&=(\eta B_1+(1-\eta)B_2)\mathrm{I}_2\\
    V2&=(\sqrt{\eta (1-\eta)}(B_1-B_2))\mathrm{I}_2\\
    V_3&=(\eta B_2+(1-\eta)B_1)\mathrm{I}_2    
\end{align}
 with 
 \begin{align}
     B_1&=G_1\Big[T_1(\frac{1}{2}-(n_{th}+\frac{1}{2}))+(n_{th}+\frac{1}{2})\Big]+\frac{G_1-1}{2}\\
     B_2&=G_2\Big[T_2(\frac{1}{2}-(n_{th}+\frac{1}{2}))+(n_{th}+\frac{1}{2})\Big]+\frac{G_2-1}{2}
 \end{align}
 where $G_1$ and $G_2$ are the amplification gains. 
 The mean vectors evolve according to the described processs as:
 \begin{align}
     d_1=0&\rightarrow \Bigg[\Bigg(\sqrt{\eta T_1}\frac{x}{\sqrt{2\langle T_1\rangle}}-\sqrt{(1-\eta) T_2}\frac{x}{\sqrt{2\langle T_1\rangle}}\Bigg) \nonumber\\
     &+ i \Bigg(\sqrt{\eta T_1}\frac{p}{\sqrt{2\langle T_1\rangle}}-\sqrt{(1-\eta) T_2}\frac{p}{\sqrt{2\langle T_1\rangle}}\Bigg)\Bigg]\nonumber\\
     d_2=x+ip&\rightarrow  \Bigg[\Bigg(\sqrt{\eta T_2}\frac{x}{\sqrt{2\langle T_2\rangle}}+\sqrt{(1-\eta) T_1}\frac{x}{\sqrt{2\langle T_2\rangle}}\Bigg) \nonumber\\
     &+ i \Bigg(\sqrt{\eta T_2}\frac{p}{\sqrt{2\langle T_2\rangle}}+\sqrt{(1-\eta) T_2}\frac{p}{\sqrt{2\langle T_2\rangle}}\Bigg) \Bigg] 
 \end{align}
A benchmark between the average fidelity in 2-diversity and 0-diversity schemes is highlighted in Fig.~\ref{fig:post_average}. Indeed, the diversity advantage depends on the average behaviour of the fading channel and on the background thermal noise. The benchmarking of the two schemes regarding both degrees of freedom of the channel is illustrated in Fig.~\ref{fig:post_contour}. 
\subsection{Pre-Amplification}
\begin{figure}
    \centering
    \includegraphics[width=1\linewidth]{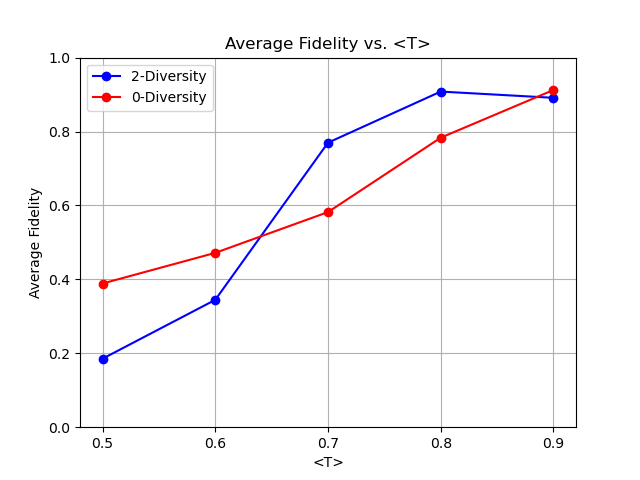}
    \caption{The average fidelity in the 2-diversity and the 0-diversity schemes as a function of the average transmittivity of the fading channel for $n_{th}=0.7$ for the thermal Gaussian noise}
    \label{fig:03}
\end{figure}

\begin{figure}
    \centering
    \includegraphics[width=1\linewidth]{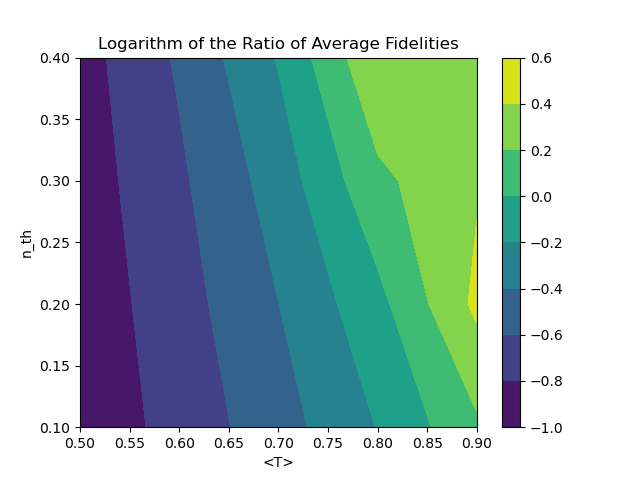}
    \caption{The log scale of the ratio between the average fidelities in the 2-diversity and the 0-diversity schemes as a function of the average transmitivitty of the fading channels and the strength of the thermal Gaussian noise.}
    \label{fig:04}
\end{figure}
We assume that CSI about the average behaviour of the fading is available to the transmitter. Accordingly, he makes a pre-amplification to overcome the effect of losses at the expense of adding an additional noise to the system.  The evolution of the covariance matrix and the mean vectors in such a setup are given by:
\begin{align}
    &V_1\oplus V_2 \rightarrow\nonumber\\
    & B(\eta)\Bigg[(K_1\oplus K_2)\Big[(A(G_1)\oplus A(G_2))(V_1\oplus V_2)(A(G_1)\oplus A(G_2))^T \nonumber\\
    &+ N_{A(G_1)}\oplus N_{A(G_2)} \Big](K_1\oplus K_2)^T +(N_1\oplus N_2) \Bigg] B(\eta)^T\\
    &d_1\oplus d_2 \rightarrow  B(\eta)(K_1\oplus K_2)  (A(G_1)\oplus A(G_2))(d_1\oplus d_2)
\end{align}
Explicitly,
\begin{align}
 V^{(2)}_{in}=\begin{pmatrix}
         \frac{\mathrm{I}_2}{2}& 0\\
         0& \frac{\mathrm{I}_2}{2}
    \end{pmatrix} \rightarrow  V^{(2)}_{out}=\begin{pmatrix}
         V1& V2\\
         V2& V3
    \end{pmatrix} 
\end{align}

\begin{align}
    V1&=(\eta B_1+(1-\eta)B_2)\mathrm{I}_2\\
    V2&=(\sqrt{\eta (1-\eta)}(B_1-B_2))\mathrm{I}_2\\
    V_3&=(\eta B_2+(1-\eta)B_1)\mathrm{I}_2    
\end{align}
 with 
 \begin{align}
     B_1&=(2G_1-1)\frac{T_1}{2}+(1-T_1)(n_{th}+\frac{1}{2})\\
     B_2&=(2G_2-1)\frac{T_2}{2}+(1-T_2)(n_{th}+\frac{1}{2})
 \end{align}
 where $G_1$ and $G_2$ are the amplification gains. 
 The mean vectors evolve according to the described processs as:
 \begin{align}
     d_1=0&\rightarrow \Bigg[\Bigg(\sqrt{\eta T_1}\frac{x}{\sqrt{2\langle T_1\rangle}}-\sqrt{(1-\eta) T_2}\frac{x}{\sqrt{2\langle T_1\rangle}}\Bigg) \nonumber \\
     &+ i \Bigg(\sqrt{\eta T_1}\frac{p}{\sqrt{2\langle T_1\rangle}}-\sqrt{(1-\eta) T_2}\frac{p}{\sqrt{2\langle T_1\rangle}}\Bigg)\Bigg]\nonumber\\
     d_2=x+ip&\rightarrow  \Bigg[\Bigg(\sqrt{\eta T_2}\frac{x}{\sqrt{2\langle T_2\rangle}}+\sqrt{(1-\eta) T_1}\frac{x}{\sqrt{2\langle T_2\rangle}}\Bigg) \nonumber\\
     &+ i \Bigg(\sqrt{\eta T_2}\frac{p}{\sqrt{2\langle T_2\rangle}}+\sqrt{(1-\eta) T_2}\frac{p}{\sqrt{2\langle T_2\rangle}}\Bigg) \Bigg] 
 \end{align}
\hspace{2cm}

\subsection{In the presence of cross talk}
\begin{figure*}
    \begin{minipage}[c] {0.3\textwidth}
        \centering
        \includegraphics[width=1\columnwidth]{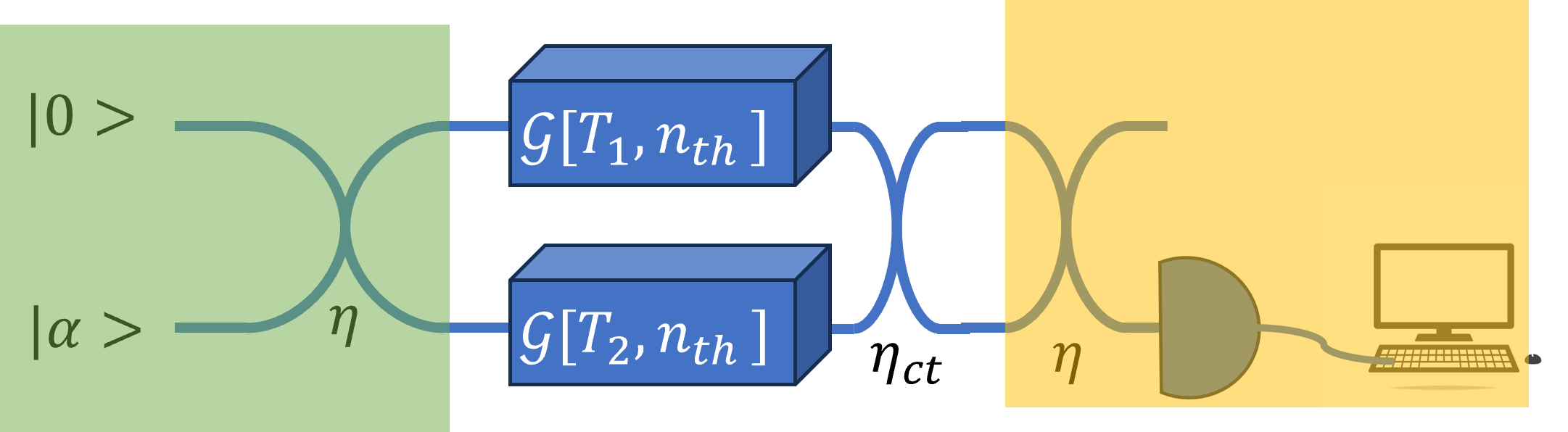}
        \subcaption{A diversity scheme with postprocessing amplification of the signal upon CSI at the receiver}
        \label{Fig:04-aa}
    \end{minipage}
    \hspace{0.02\textwidth}
   \begin{minipage}[c] {0.3\textwidth}
        \centering
        \includegraphics[width=1\columnwidth]{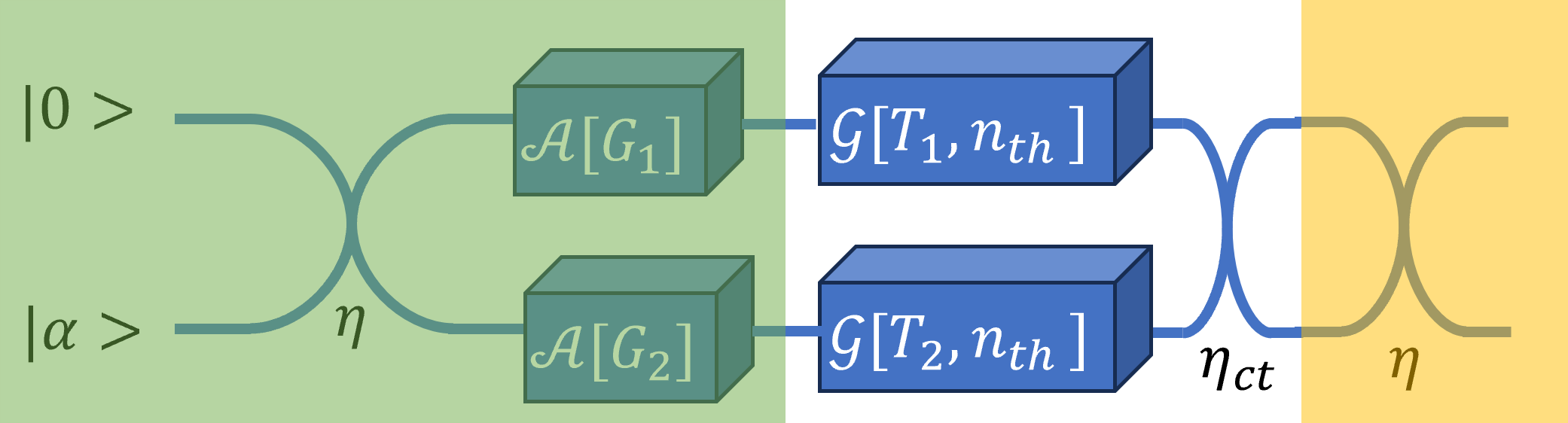}
        \subcaption{A diversity scheme with postamplification of the signal upon CSI at the receiver}
        \label{Fig:04-bb}
    \end{minipage}
    \hspace{0.02\textwidth}
      \begin{minipage}[c] {0.3\textwidth}
        \centering
        \includegraphics[width=1\columnwidth]{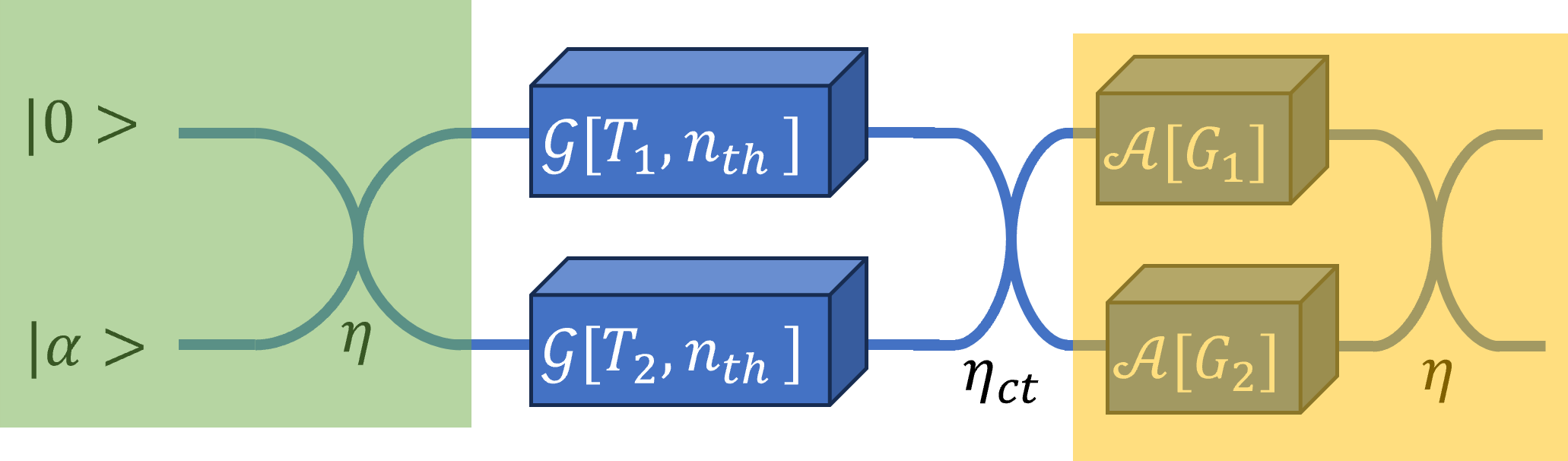}
        \subcaption{A diversity scheme with preamplification of the signal with CSI at the transmitter.}
        \label{Fig:04-cc}
    \end{minipage}
    \caption{A diversity scheme with different amplification techniques in the presence of cross talk with green and yellow shadowed boxes refering to the transmitter and the receiver respectively.}
    \label{fig:diversity_cross_talk}
    \hrulefill
\end{figure*}
\begin{figure}[t]
    \centering
    \includegraphics[width=1\linewidth]{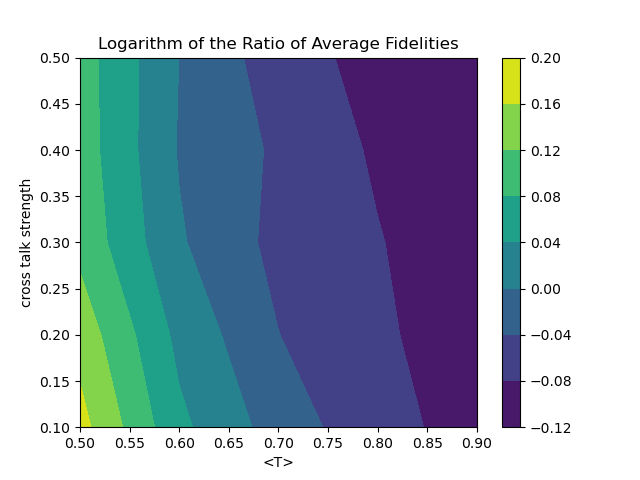}
    \caption{Average Fidelity ratio in the presence of cross-talk in the postprecessing amplification scenario. The strength of the thermal noise is kept fixed at $n_{th}=0.2$}
    \label{fig:05}
\end{figure}

\begin{figure}[t]
    \centering
\includegraphics[width=1\columnwidth]{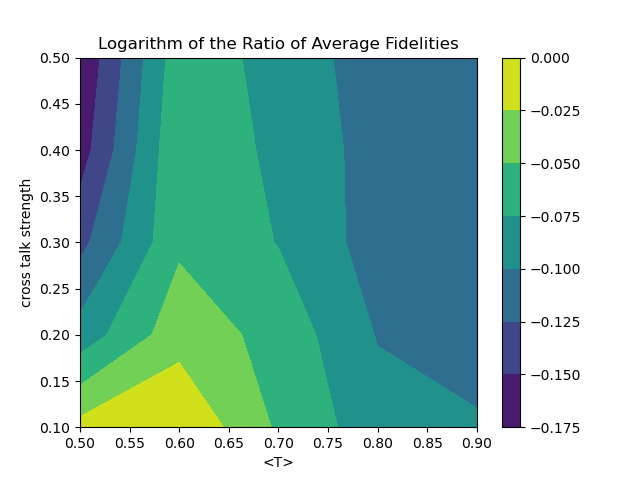}
    \caption{Average Fidelity ratio in the presence of cross-talk in the pre-amplification scenario. The strength of the thermal noise is kept fixed at $n_{th}=0.2$}
    \label{fig:post_}
\end{figure}

\begin{figure}[t]
    \centering
\includegraphics[width=1\columnwidth]{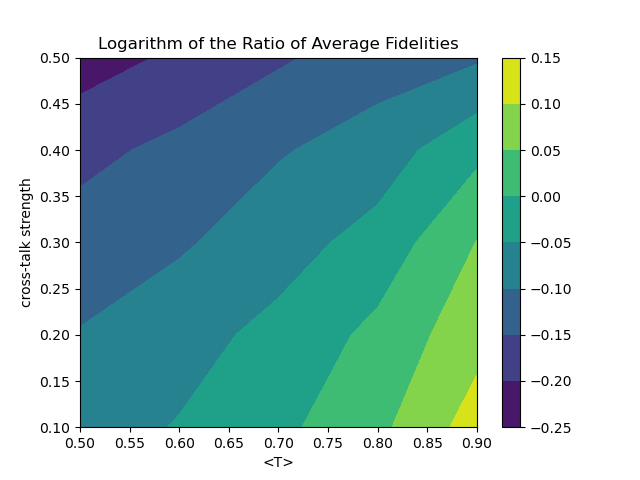}
    \caption{Average Fidelity ratio in the presence of cross-talk in the pre-amplification scenario. The strength of the thermal noise is kept fixed at $n_{th}=0.2$}
    \label{fig:06}
\end{figure}

We modalize crosstalk in CV quantum communications as a beam splitter with stochastic transmitivitty paramter as discussed in Sec.~\ref{sec:02}.
In the presence of crosstalk the covariances and the mean vector of the system undergo the global evolution given in Eq~.\ref{postproc},\ref{post},\ref{pre}. Therein, $B(\eta_{ct})$ denotes the crosstalk noise between the two channels with $\eta_{ct}$ being the crosstalk parameters reflecting the strength of the interference between the transmitted signals. The simulations are presented in Fig.~\ref{fig:05}, Fig.~\ref{fig:post_} and Fig.~\ref{fig:06}  respectivelycfor postprocessing amplification, postamplification and preamplification. 
We note that in the presence of cross-talk, the advantage brought by the 2-diversity scheme is reduced for postprocessing, postamplification and pre-amplification respectively, but an advantage is still observed at strong fading and strong cross-talk for postprocessing and post-amplification. In the contrary, no advantage is witnessed for postamplification in the presence of crosstalk. This reflects the fact that in postamplification, the noise in the transmission paths, be it loss or thermal, is amplified along with the signal, leading to a noisy state at the output.

\begin{figure*}
\begin{align}
    V_1 \oplus V_2 &\rightarrow B(\eta) B(\eta_{ct}) \Big[ (K_1 \oplus K_2) (V_1 \oplus V_2) (K_1 \oplus K_2)^T + (N_1 \oplus N_2) \Big] B(\eta_{ct})^T B(\eta)^T \nonumber\\
    d_1 \oplus d_2 &\rightarrow G \cdot B(\eta) B(\eta_{ct}) (K_1 \oplus K_2) (d_1 \oplus d_2)
    \label{postproc}
\end{align}
\vspace*{4pt}
\begin{align}
    V_1 \oplus V_2 &\rightarrow B(\eta) B(\eta_{ct}) \Bigg[ (A(G_1) \oplus A(G_2)) \Big[ (K_1 \oplus K_2) (V_1 \oplus V_2) (K_1 \oplus K_2)^T \nonumber \\
    &+ (N_1 \oplus N_2) \Big] (A(G_1) \oplus A(G_2))^T + N_{A(G_1)} \oplus N_{A(G_2)} \Bigg] B(\eta_{ct})^T B(\eta)^T \nonumber\\
    d_1 \oplus d_2 &\rightarrow  B(\eta) B(\eta_{ct}) (K_1 \oplus K_2) (A(G_1) \oplus A(G_2)) (d_1 \oplus d_2)
    \label{post}
\end{align}
\vspace*{4pt}
\begin{align}
    V_1 \oplus V_2 &\rightarrow B(\eta) B(\eta_{ct}) \Bigg[ (K_1 \oplus K_2) \Big[ (A(G_1) \oplus A(G_2)) (V_1 \oplus V_2) (A(G_1) \oplus A(G_2))^T \nonumber \\
    &+ N_{A(G_1)} \oplus N_{A(G_2)} \Big] (K_1 \oplus K_2)^T + (N_1 \oplus N_2) \Bigg] B(\eta_{ct})^T B(\eta)^T \nonumber\\
    d_1 \oplus d_2 &\rightarrow  B(\eta) B(\eta_{ct}) (A(G_1) \oplus A(G_2)) (K_1 \oplus K_2) (d_1 \oplus d_2)
    \label{pre}
\end{align}
\end{figure*}

\subsection{Secret key rate}
\begin{figure*}

    \begin{minipage}[c] {0.49\textwidth}
        \centering
    \includegraphics[width=0.9\columnwidth]{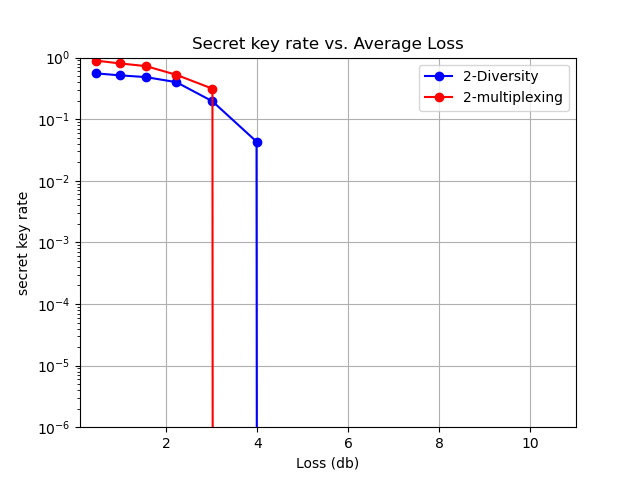}
    \subcaption{A benchmark between the secret key rate in the 2-diversity and the 2-multiplexing cases with heterodyne detection for perfect reconciliation efficiency and  variance $V_{in}=10$}
    \label{fig:skr}
    \end{minipage}
    \hspace{0.02\textwidth}
    \begin{minipage}[c] {0.49\textwidth}
         \centering
    \includegraphics[width=0.9\columnwidth]{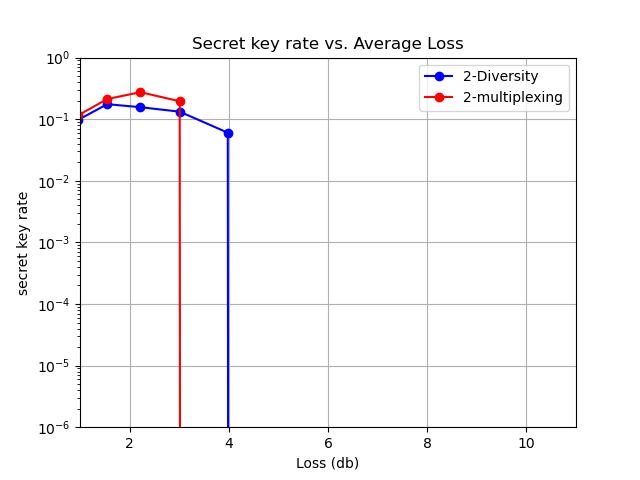}
    \subcaption{A benchmark between the secret key rate in the 2-diversity and the 2-multiplexing cases with homodyne detection for perfect reconcilliation efficiency and  variance $V_{in}=5$}
    \label{fig:skr_homodyne}   
    \end{minipage}
    \vspace{6pt}

    \caption{CV QKD secret key rate}
    \label{fig:multiplexing-diversity}
\end{figure*}
\begin{figure*}
    \begin{minipage}[c] {0.49\textwidth}
        \centering
    \includegraphics[width=0.9\columnwidth]{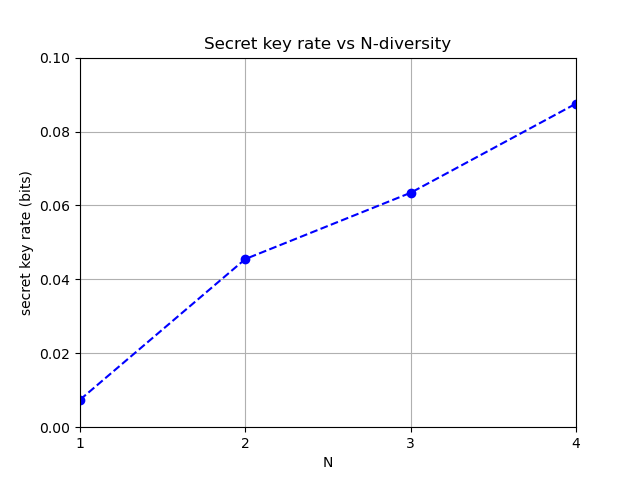}
    \subcaption{The average secret key rate with heterodyne detection as a function of the number of used channels for diversity for variance $V_{in}=10$ and $4$db loss}
    \label{fig:skr1}
    \end{minipage}
    \hspace{0.02\textwidth}
    \begin{minipage}[c] {0.49\textwidth}
         \centering
    \includegraphics[width=0.9\columnwidth]{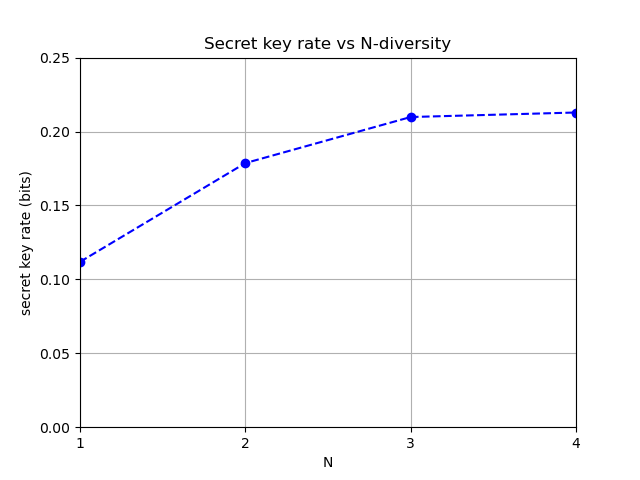}
    \subcaption{The average secret key rate with homodyne detection as a function of the number of used channels for diversity for variance $V_{in}=5$ and $2$db loss}
    \label{fig:skr_homodyne1}   
    \end{minipage}
    \vspace{6pt}

    \caption{CV QKD secret key rate as function of N, the number of used channels for diversity}
    \label{fig:N_diversity}
\end{figure*}
A direct consequence of the fidelity increase brought by the diversity scheme is the possibility to decrease the probability of error of decoding a given state, resulting in higher secret key rate in CV-QKD. To study this, We consider the same prepare and measure scenario used until now, where Alice, the transmitter, prepares a coherent state with respect to some specific modulation, and sends it to Bob, the receiver, who performs homodyne or heterodyne measurement. Indeed, the scheme is equivalent to an entanglement based scheme where  Alice and Bob share a two mode squeezed entangled state, hence the same security. 
The mutual information $I(A:B)$ in this case is given by (see Appendix.~\ref{app:mutual info} for derivation):
\begin{equation}
    I(A:B)=\frac{\nu}{2} \log_2\Bigg[1+\frac{\nu (\frac{T1+T2}{2})(V_{in}-1)}{1+\frac{1}{\nu}(1-\frac{T1+T2}{2})(2 n_{th}-1)}\Bigg]
\end{equation}
with $\nu=1$ for homodyne detection and $\nu=2$ for heterodyne dtection. 
Moreover, to obtain the Holevo information, the reduced covariance matrices on the environment $V_E$ and the covariance matrix of the environment conditioned on the measurement outcomes of Bob $V_{E|B}$ and their respective symplectic eigenvalues are needed. 
The environment covariance matrix $V_{E}$ is given by (see Appendix.~\ref{app:covariance env} for derivation): 
\begin{equation}
    \label{covariance_env}V_{E}=\begin{pmatrix}
        a\mathrm{I}_2 & c\sigma_z\\
        c\sigma_z & b\mathrm{I}_2
    \end{pmatrix}
\end{equation}
with:
\begin{align}
    a&=\Big(1-\frac{T1+T2}{2}\Big)V_{in}+\frac{(T1+T2)}{2}\Big(n_{th}+\frac{1}{2}\Big) \nonumber\\
    b&= n_{th}+\frac{1}{2}\nonumber\\
    c&=\sqrt{\frac{(T1+T2)}{2}\Big((n_{th}+\frac{1}{2})^2\Big)-1}
\end{align}
The symplectic eigenvalues of this matrix are given by:
\begin{equation}
    \nu_\pm=\frac{1}{2}(z\pm (b-a))
\end{equation}
where:
\begin{equation}
    z=\sqrt{(a+b)^2-4c^2}
\end{equation}

Differently, the conditional covariance matrix of the environment $V_{E|B}$ depends on the type of measurement. Due to the entropic equality:
\begin{equation}
    S(E)-S(E|B)=S(AB)-S(A|B)
\end{equation}
we focus on $V_{A|B}$ instead of $V_{E|B}$
We distinguish two cases, heterodyne detection and homodyne detection. For heterodyne detection $V_{A|B}$ is given by:
\begin{equation}
    V_{E|B}=\Big(k-\frac{g^2}{h+1}\Big)\mathrm{I}_2
\end{equation}
with:
\begin{align}
    k&=V_{in} \nonumber\\
    h&=\Big(\frac{T1+T2}{2} \Big)(V_{in}-1)+(n_{th}-\frac{1}{2})\Big(1-\frac{T1+T2}{2}\Big)+1\nonumber\\
    g&=\sqrt{\Big(\frac{T1+T2}{2} \Big)(V_{in}^2-1)} 
\end{align}
with symplectic eigenvalues:
\begin{equation}
    \nu_{ht}=k-\frac{g^2}{h+1}
\end{equation}

For homodyne detection, $V_{A|B}$ is given by:
\begin{equation}
    V_{A|B}=\begin{pmatrix}
        k-\frac{g^2}{h} & 0\\
        0 & k
    \end{pmatrix}
\end{equation}
with symplectic eigenvalue:
\begin{equation}
    \nu_{hm}=\sqrt{k\Big(k-\frac{g^2}{h}\Big)}
\end{equation}
The simulations of the average secret key rate in the presence of fading channels are reported in Fig.~\ref{fig:skr}. To see the advantage of diversity in CV-QKD, we consider the 2-multiplexing case, where the QKD channel is bi-multiplexed without considering any interference effects or cross talk between the two channels. This is highlighted in Fig.~\ref{fig:multiplexing-diversity}.  We clearly notice a diversity gain for the secret key generated between Alice and Bob when the signal is transmitted through different paths and recombined at Bob's side both using heterodyne and homodyne detections at high losses. The scheme has extended for multiple streams in parallel used for diversity and the result on the average secret key rate is presented in Fig.~\ref{fig:N_diversity}where we see that increasing the channels used for diversity increases the average secret key rate. These results result imply a  diversity advantage when multiplexing fails to increase the key rate in practical quantum communications which can be harnessed to push the CV-QKD systems to their limits.

\section{Conclusions}
\label{sec:04}
In this paper, we have analyzed the performance of continuous-variable (CV) quantum communication systems under realistic channel conditions, including Gaussian lossy channels, fading, and crosstalk. We introduced diversity schemes to enhance the robustness of these systems against such impairments. By modeling the transmittivity of the channel as a log-normal distribution, we have accounted for the stochastic nature of fading, which is a significant factor in real-world quantum communications.

Our results demonstrate that diversity schemes, particularly those involving multiple transmission paths, provide substantial improvements in terms of fidelity over single-channel transmission. These improvements are most pronounced under conditions of strong fading and high thermal background noise. We also investigated the impact of crosstalk, showing that while it reduces the performance gain, diversity schemes still offer noticeable advantages in adverse channel conditions. 

For practical application, we considered the advantage of diversity in CV-QKD setups. Our results demonstrated a diversity gain in the high loss scenarios with respect to multiplexing. This result implies a diversity advantage considerations in practical quantum communications, and urges further research towards the understanding of the diversity-multiplexing tradefoff in the presence of signal interference and crosstalk between different channels. 

\appendices

\section{The mutual information $I(A:B)$}
\label{app:mutual info}
Assuming the modulation variance at the sender Alice being $V_{in}$, then after transmission of the state through a bosonic Gaussian channel the covariance matrix undergoes the following transformation:
\begin{equation}
    V_{in}\rightarrow T V_{in} + (n_{th}+\frac{1}{2})\mathrm{I}_2
\end{equation}
as specified in Eq.~\ref{gaussian channel}. In a diversity scheme setup, and in the presence of two parallel channels with similar combining of Figure.~\ref{fig:diversity} after discarding local oscilator state, the covariance matrix evolves as:
\begin{equation}
    V_{in}\rightarrow \frac{T_1+ T_2}{2} V_{in}+ (1-\frac{T_1+ T_2}{2})(n_{th}+\frac{1}{2})\mathrm{I}_2
    \label{qkd-homodyne}
\end{equation}
for homodyne detection, and
\begin{equation}
    V_{in}\rightarrow \frac{1}{2}\Big[\frac{T_1+ T_2}{2} V_{in}+ (1-\frac{T_1+ T_2}{2})(n_{th}+\frac{1}{2})\mathrm{I}_2\Big]+\frac{1}{4}\mathrm{I}_2
    \label{qkd-heterodyne}
\end{equation}
for heterodyne detection. 
The mutual information is defined as 
\begin{equation}
    I(A:B)=\frac{\nu}{2}\log_2(1+SNR)
\end{equation}
with $SNR$ being the signal-to-noise ratio and $\nu$ is a factor identical to $2$ for heterodyne detection and $1$ for homodyne detection. 
From Eq.~\ref{qkd-homodyne} and Eq.~\ref{qkd-heterodyne} we can deduce the $SNR$ for both homodyne and heterodyne detections and  it is given respectively by:
\begin{equation}
    SNR= \frac{\nu (\frac{T1+T2}{2})(V_{in}-1)}{1+\frac{1}{\nu}(1-\frac{T1+T2}{2})(2 n_{th}-1)}
\end{equation}
therefore the mutual information is given by:
\begin{equation}
    I(A:B)=\frac{\nu}{2} \log_2\Bigg[1+\frac{\nu (\frac{T1+T2}{2})(V_{in}-1)}{1+\frac{1}{\nu}(1-\frac{T1+T2}{2})(2 n_{th}-1)}\Bigg]
\end{equation}

\section{The covariance matrix of the environment}
\label{app:covariance env}
\begin{figure}
    \centering
    \includegraphics[width=0.9\columnwidth]{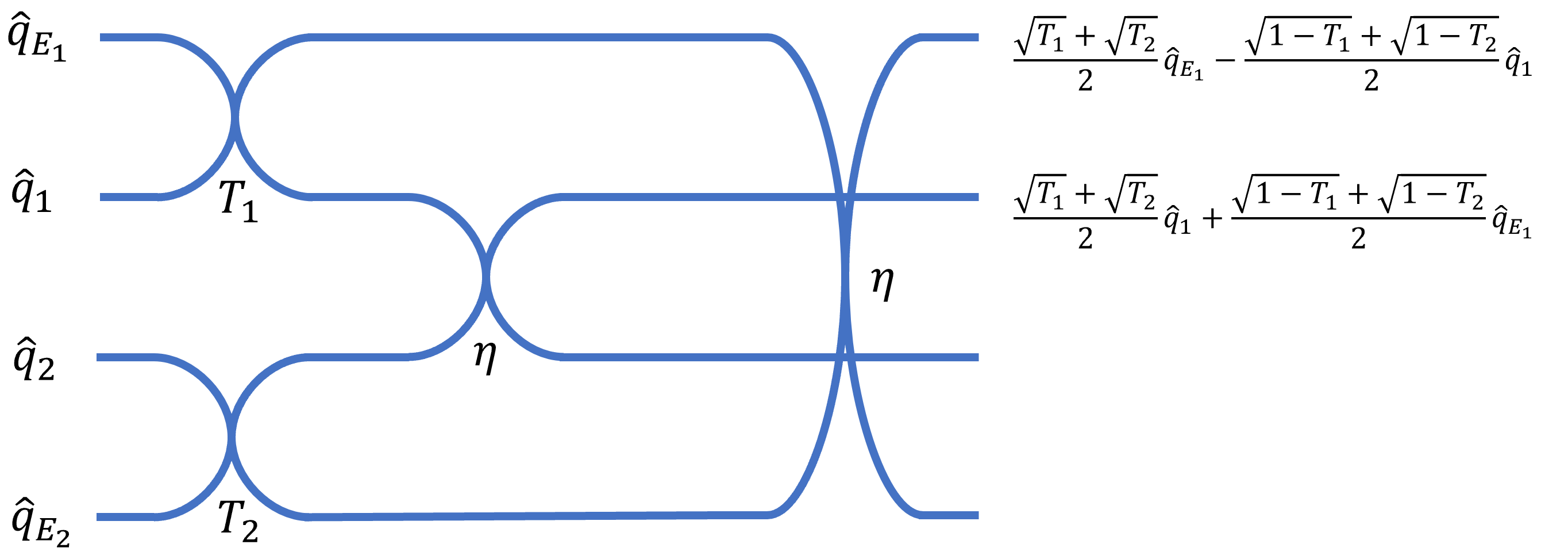}
    \caption{A diagram showing the evolution of the quadratures of different modes, where $\hat{q}_1$ and $\hat{q}_2$ stand for the signal modes transmitted through the Gaussian channels. The gaussian channels are  respresented by  beam splitter interactions with tranmitivities $T_1$ and $T_2$ respectively, with $\hat{q}_{E_1}$ and $\hat{q}_{E_2}$  standing for the attacker modes. The signal modes and the attacker modes are combined finally with $50:50$ beam splitters. One of the signal modes is discarded as well as one of the attacker modes. }
    \label{qkd_environment}
\end{figure}

To understand the derivation  of the covariance matrix of the environment $V_E$ in Eq.~\ref{covariance_env} we follow the diagram in Fig.~\ref{qkd_environment} reflecting the evolution of the quadratures in the presence of an environment which is controlled by the attacker. The quadratures $\hat{q}_1$ and $\hat{q}_2$ stand for the signal modes which are transmitted through two parallel Gaussian channels for the 2-diversity scheme. The two Gaussian channels are explicitly represented by beam splitter interactions of the signal modes with the corresponding attacker modes, $\hat{q}_{E_1}$ and $\hat{q}_{E_2}$ with transmittivities $T_1$ and $T_2$ respectively. The signal modes are combined at the receiver by a $50:50$ beam splitter and the same for the attacker modes. At the end of the process, one of the signal modes are discarded as well as one of the attacker modes, reflecting a cloning attack. The quadratures of the remaining signal and environment mode are given by:
\begin{align}
    \hat{q}_B&=\frac{\sqrt{T_1}+\sqrt{T_2}}{2}\hat{q}_1+\frac{\sqrt{1-T_1}+\sqrt{1-T_2}}{2}\hat{q}_{E_1}\\
    \hat{q}_{E_o}&=\frac{\sqrt{T_1}+\sqrt{T_2}}{2}\hat{q}_{E_1}-\frac{\sqrt{1-T_1}+\sqrt{1-T_2}}{2}\hat{q}_{1}
\end{align}
By setting $\langle \hat{q}_{E_1}^2\rangle= \langle\hat{p}_{E_1}^2\rangle=(n_{th}+\frac{1}{2})$ and $\langle\hat{q}_{1}^2\rangle=\langle\hat{p}_{1}^2\rangle=V_{in}$, the covariances and the cross covariances  of an entangled attacker are given by:
\begin{align}
  \langle \hat{q}_{E_o}^2 \rangle &= \langle \hat{p}_{E_o}^2 \rangle=\frac{T_1+T_2}{2} (n_{th}+\frac{1}{2}) + \Big(1-\frac{T_1+T_2}{2}\Big)V_{in}\nonumber\\
  \langle \hat{q}_{E_o}\hat{q}_{E}\rangle&= -\langle \hat{p}_{E_o}\hat{p}_{E}\rangle=\frac{\sqrt{T_1}+\sqrt{T_2}}{2} \langle \hat{q}_{E_1}\hat{q}_{E}\rangle \nonumber\\
  &=\frac{\sqrt{T_1}+\sqrt{T_2}}{2}\sqrt{(n_{th}+\frac{1}{2})^2-1}\nonumber\\
  \langle\hat{q}_{E}^2\rangle &= \langle\hat{p}_{E}^2\rangle = (n_{th}+\frac{1}{2})
\end{align}
with $\hat{q}_{E}$ and $\hat{p}_{E}$ being the quadratures of the entangling attacker mode. As a result, the covariance matrix of the environment is:
\begin{equation}
    V_{E}=\begin{pmatrix}
        a\mathrm{I}_2 & c\sigma_z\\
        c\sigma_z & b\mathrm{I}_2
    \end{pmatrix}
\end{equation}
with:
\begin{align}
    a&=\Big(1-\frac{T1+T2}{2}\Big)V_{in}+\frac{(T1+T2)}{2}\Big(n_{th}+\frac{1}{2}\Big) \nonumber\\
    b&= n_{th}+\frac{1}{2}\nonumber\\
    c&=\sqrt{\frac{(T1+T2)}{2}\Big((n_{th}+\frac{1}{2})^2-1\Big)}
\end{align}

\bibliographystyle{IEEEtran}
\bibliography{ref}
\end{document}